\documentclass[twoside]{dis07}
\usepackage[latin1]{inputenc}
\usepackage[dvips]{graphicx,epsfig,color}
\usepackage{subfigure}
\usepackage{wrapfig,rotating}
\usepackage{amssymb,amsmath,array}

\newcommand {\snn}      {$\sqrt{s_{_{\rm NN}}}$}

\newcommand {\pt}       {p_{\rm T}}

\begin{document}
\title{Recent heavy flavor results from STAR}

\author{Andr\'e Mischke\thanks{The author is grateful for the support by the Netherlands Organisation for Scientific Research (NWO).}
\hspace{.02cm} for the STAR Collaboration\footnote{For the full author list, see~\cite{Paper:npe}.
}
%
\vspace{.3cm}\\
Institute for Subatomic Physics, Utrecht University, \\ 
Princetonplein 5, 3584 CC Utrecht, \\ The Netherlands. \\
\emph{E-mail: a.mischke@phys.uu.nl} \\
}

\maketitle

\begin{abstract}
We report on recent heavy flavor measurements from the STAR experiment 
at RHIC\cite{url}. 
The measured charm cross section in heavy-ion collisions scales 
with the number of binary collisions, which is an indication for exclusive charm 
production in the initial state of the collision. The observed strong suppression 
of non-photonic electrons at high $\pt$ in Au+Au collisions together with the 
azimuthal correlation measurements in p+p collisions imply a suppression of
$B$ production in heavy-ion collisions. We also present 
recent measurements of the $\Upsilon$ cross section in p+p collisions.\\
\end{abstract}

\section{Introduction}
The fundamental theory of strong interactions, Quantum Chromodynamics 
(QCD), predicts a phase transition from hadronic matter to a system of 
deconfined quarks and gluons, the Quark Gluon Plasma, if the surrounding 
temperature exceeds a critical value. The goal of heavy-ion physics is to 
produce such a deconfined QCD state and to study its properties under 
controlled laboratory conditions. 
The accelerator with the current highest collision energy for atomic nuclei 
is the Relativistic Heavy-Ion Collider (RHIC) at Brookhaven National Laboratory.
Current results from the RHIC experiments have given compelling evidences 
that the produced medium is indeed a plasma of quarks and gluons, but it 
behaves like a ``perfect'' fluid rather than an ideal 
gas~\cite{Star:3years, Peter:overview}. 
One of the intriguing results is the strong modification of the jet structure 
inside the created medium. Theoretical model calculations that attribute the 
jet attenuation to the energy loss of partons traversing through the medium 
have successfully described the present data.

The study of heavy flavor (charm, bottom) production in heavy-ion collisions 
provides key tests of the parton energy loss mechanisms and offers important 
information on the properties of the produced medium. Due to their large mass, 
heavy quarks are expected to be primarily produced in 
the early stage of the collision and, therefore, probe the complete space-time 
evolution of the medium. Theoretical models predicted that heavy quarks should 
experience smaller energy loss than light quarks while propagating through the 
QCD medium due to the suppression of small angle gluon 
radiation (dead-cone effect)~\cite{Deadcone}.

\begin{figure}[htb]
\vspace*{-.8cm}
\hspace*{0.3cm}
\subfigure{\includegraphics[width=0.46\textwidth]{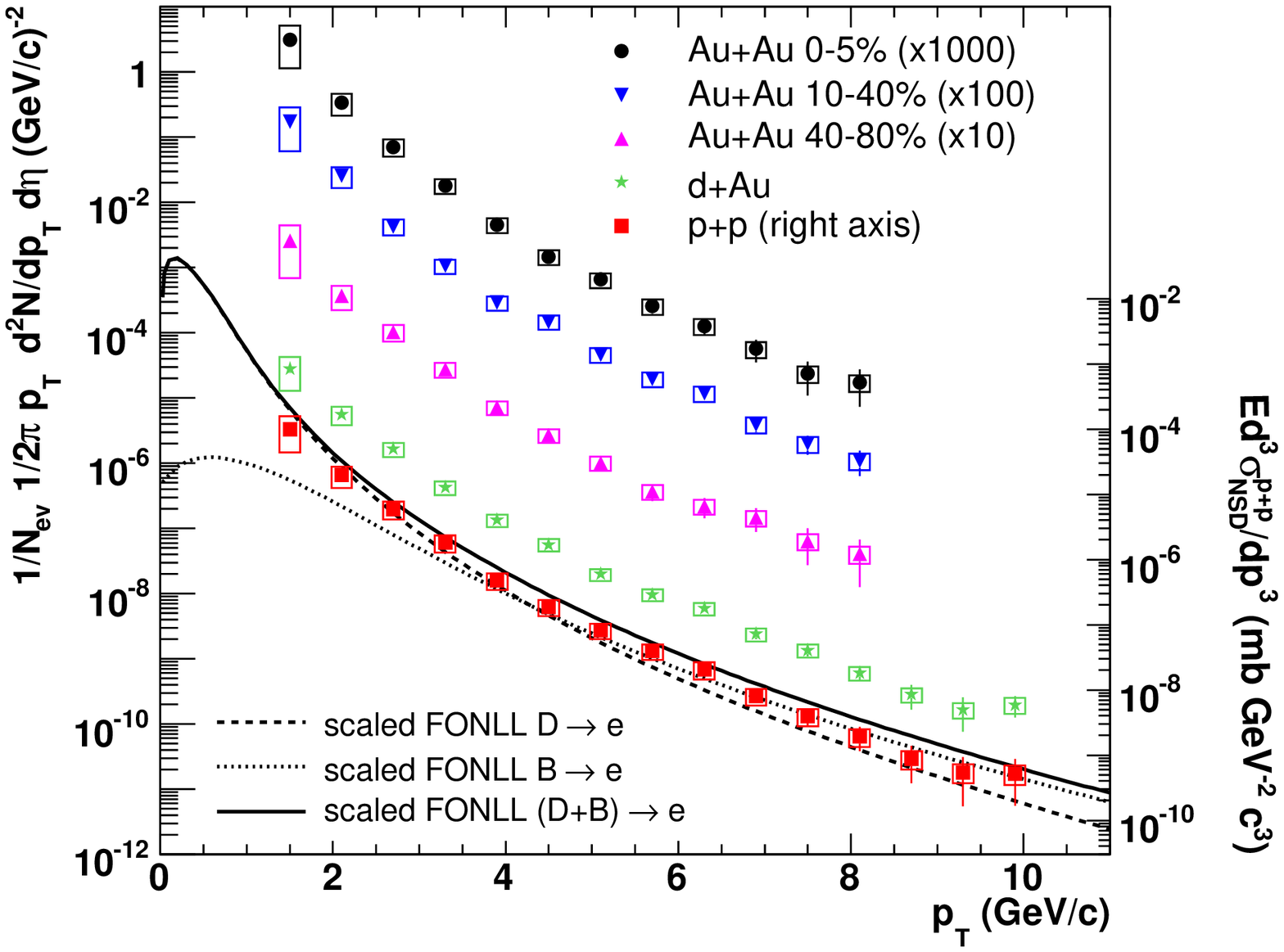}}
\hspace*{0.4cm}
\subfigure{\includegraphics[width=0.42\textwidth]{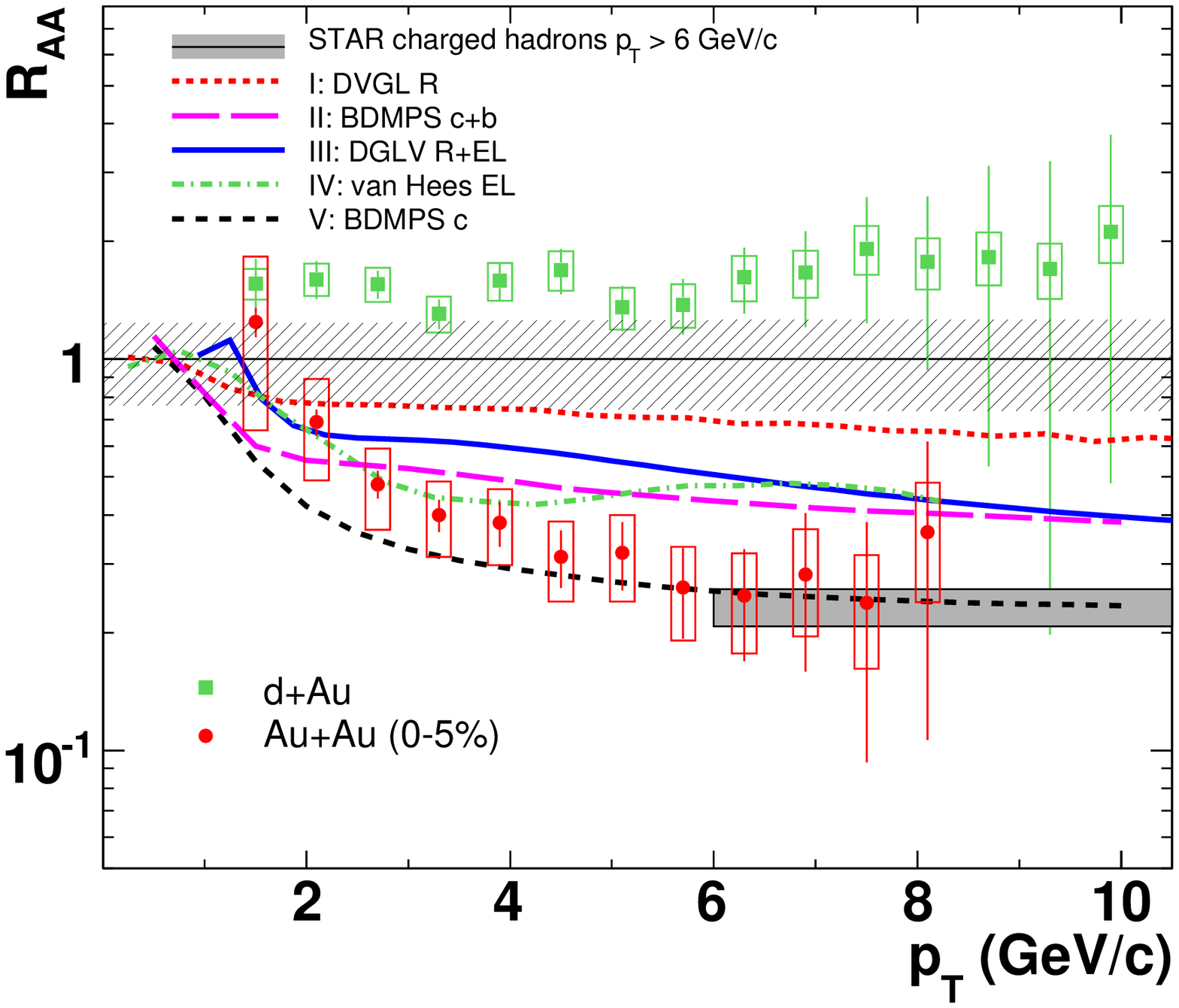}}
\vspace*{-0.45cm}
\caption[]{\label{fig:1}
Left panel: Non-photonic electron spectra in p+p, d+Au and, for different centralities, 
in Au+Au collisions at \snn = 200 GeV. The curves are pQCD predictions scaled 
by 5.5 for p+p collisions. The right axis gives the cross section for the p+p spectrum. 
Right panel: Nuclear modification factor $R_{AA}$ for d+Au and Au+Au collisions. 
The curves correspond to different model predictions as indicated in the figure.
The error bars (boxes) indicates the statistical (uncorrelated systematic) errors. 
The dashed box illustrates the overall normalization uncertainty.
}
\end{figure}

\section{Recent results}
The charm cross section at mid-rapidity is determined from measurements 
of open charm mesons and from the reconstruction of heavy flavor 
semi-leptonic decays via muon and electron measurements. 
These three measurements, which are performed by the STAR Time Projection 
Chamber (TPC) and Time-of-Flight (ToF) detector, cover 95$\%$ of the cross 
section. $D^{0}$ mesons are reconstructed in the hadronic decay channel 
$D^{0} \rightarrow K^{-} \pi^{+}$ by calculating the invariant mass of all 
oppositely charged TPC track combinations~\cite{D0:dAu, D0:AuAu}.
The decay particles are identified using the specific energy loss ($dE/dx$)
measured in the TPC.
Muons at low $\pt$ ($<$ 250 MeV/c) are identified by the combination of the 
$m^{2}$ measurement in the ToF detector and the specific energy loss 
($dE/dx$) measurement in the TPC.
A cut on the distance of closest approach is used to separate the prompt 
from decay muons.
The non-photonic electrons are obtained by combining the $dE/dx$ and ToF 
($|1/\beta-1| <$ 0.03) measurement.
A description of the determination of the photonic electron background, 
the applied corrections and the procedure to calculate the charm cross 
section from the $\pt$ spectra of $D^{0}$, $\mu$ and $e$ is given in~\cite{D0:dAu}.
The obtained cross section is found to be 
$\sigma^{NN}_{cc} = 1.40\pm0.11\pm0.39$ mb in the 12$\%$ most central 
Au+Au collisions.
NLO calculations predict a factor of $\approx5$ smaller value for the cross 
section. 
More precise measurements are required in order to understand this discrepancy.
Moreover, the cross section at mid-rapidity shows binary collision scaling 
which is an indication for charm production exclusive in the initial state 
of the collisions~\cite{Zhang}. 
Hence, there is no room for thermal charm production in the medium. \\

\begin{figure}[htb]
\vspace*{-.8cm}
\hspace*{1.1cm}
\subfigure{\includegraphics[width=0.37\textwidth]{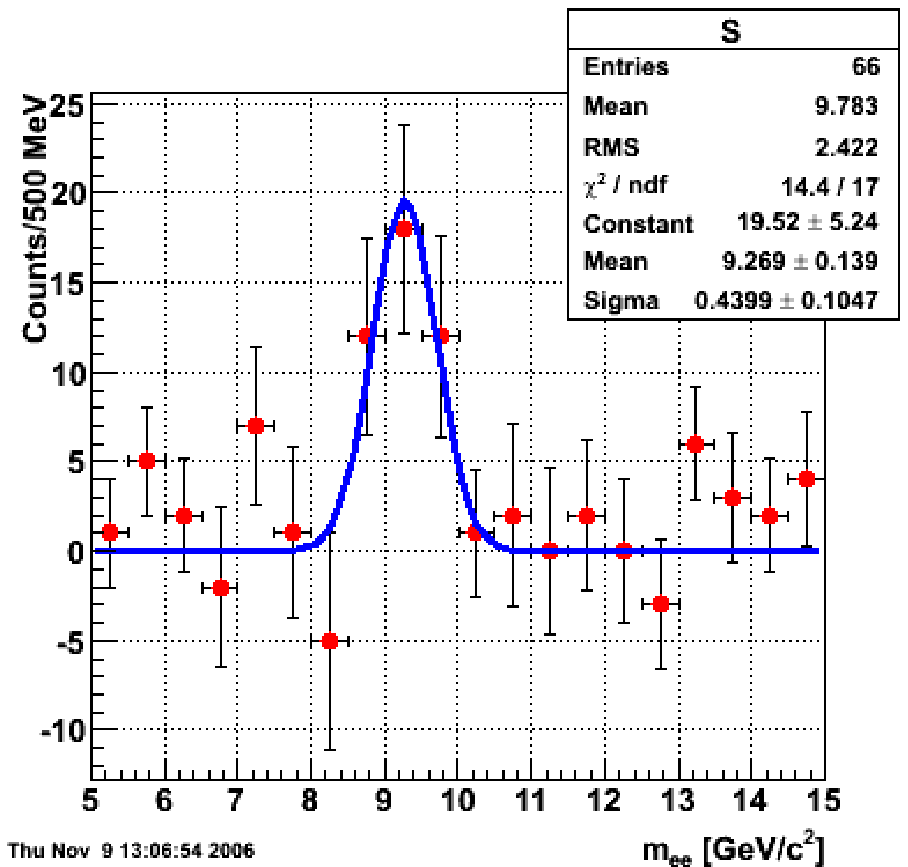}}
\hspace*{2.cm}
\subfigure{\includegraphics[height=0.37\textwidth]{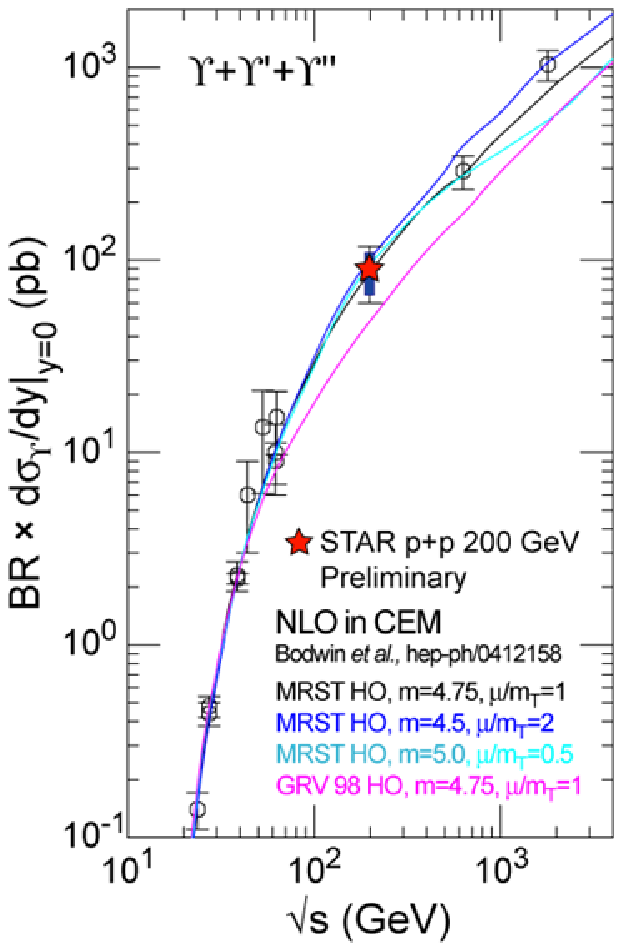}}
\vspace*{-0.4cm}
\caption[]{\label{fig:2}
Left panel: Background subtracted di-electron invariant mass distribution in 
p+p collisions at \snn = 200 GeV.
The blue line indicates a Gaussian fit to the data.
Right panel: Excitation function of the $\Upsilon$ cross section. 
The data is compared to previous measurements and NLO calculations.
The curves of the NLO calculations are scaled by a factor of 1.44 to 
account for the excited states.}
\end{figure}

Electrons at higher $\pt$ ($>$ 4 GeV/c) are identified by a combined measurement 
using the TPC and the Electromagnetic Calorimeter (EMC). 
The analysis details are described in~\cite{Paper:npe}.
Fig.~\ref{fig:1} (left) shows the $\pt$ spectrum of non-photonic electrons in p+p, 
d+Au and, for different centralities, in Au+Au collisions at \snn = 200 GeV, which 
are measured up to 10 GeV/c.
A pQCD calculation for heavy quarks production in p+p collisions~\cite{Cacciari}, 
indicated in Fig.~\ref{fig:1} (left) by the solid line, describes the overall shape of 
the $\pt$ distribution but it has the same scaling discrepancy as observed for the
charm cross section measurement.
Nuclear effects are usually quantified in the nuclear modification factor $R_{AA}$ 
where the yield in Au+Au is divided by the yield in p+p scaled by the number of 
binary collisions.
The non-photonic electron yield exhibits an unexpectedly large suppression in 
central Au+Au collisions at high $\pt$, suggesting substantial energy loss of 
heavy quarks in the produced medium (cf. Fig.~\ref{fig:1}, right). 
The suppression factor has a similar value as observed for light quark hadrons 
in central Au+Au collisions, indicated by the grey box in the figure.
The data is compared to different energy loss 
models~\cite{T:Magdalena, T:Armesto, T:Wicks, T:Hess} which vary essentially
in the interaction processes and energy loss mechanisms taken into account.
As indicated in Fig.~\ref{fig:1} (right), all models underpredict the measured 
suppression factor at high $\pt$.
It has been shown that the data is described reasonably well if the models assume 
electrons from $D$ decays only.
Therefore, the observed discrepancy could indicate that the $B$ dominance over 
$D$ mesons starts at higher $\pt$.
A possible scenario for $B$ meson suppression invokes collisional dissociation 
in the medium~\cite{T:Vitev}. \\ 

To verify the $B$ production dominance at higher $\pt$ one has to disentangle 
the $D$ and $B$ contribution to the non-photonic electron spectrum experimentally.
Recent results on measurements in p+p collisions of the azimuthal angular 
correlations between electrons (from heavy-flavor decays) associated with charged 
hadrons have shown that the relative $B$ contribution, 
$B/(B+D)$, is about 40$\%$ at $\pt$ = 5 GeV/c~\cite{Xiao}. 
The measured $\pt$ dependence of the relative $B$ contribution can be 
used to verify the input parameters for most of the energy loss models.
First results on a different approach show the proof of principle to disentangle 
the $D$ and $B$ contributions to the non-photonic electrons using 
electron-$D^0$ meson azimuthal correlations~\cite{Mischke:eD0}. \\

The suppression of heavy quarkonium states provides an essential tool to 
study the temperature of the medium. The large acceptance ($|\eta|<1$ and 
$0<\phi<2\pi$) of the STAR TPC 
and EMC allows the measurement of $\Upsilon$ production at mid-rapidity.
The $\Upsilon$ is reconstructed through the $\Upsilon \rightarrow e^{+} e^{-}$ 
decay channel.
Both detectors have very good electron identification capabilities and 
allow the combined measurement of the momentum (TPC) and the 
energy (EMC) of the decay electrons.
Details of the data analysis can be found in~\cite{Mauro:upsilon}.
Due to the finite momentum resolution of the TPC, individual $\Upsilon$
states, 1S, 2S and 3S, can not be resolved.
The EMC serves as a trigger for high momentum electrons utilizing 
two dedicated trigger settings~\cite{Pibero:quarkonia}.
The presented data are from the 2006 Run with an integrated luminosity of
$\mathcal{L} = 5.6~\mathrm{pb}^{-1}$.  
The invariant mass distribution of unlike-sign electron pairs is shown in 
Fig.~\ref{fig:2} (left).
The peak width is consistent with the expected mass resolution.
The corrections applied to the raw yield are discussed in~\cite{Mauro:upsilon}.
The obtained production cross-section of 
$BR_{ee} \times \frac{d\sigma}{dy}_{y=0} = 91\pm28(stat.)\pm22(sys.)$ pb 
follows the world data trend (cf. Fig.~\ref{fig:2}, right). 
Within uncertainties, the data shows very good agreement with NLO calculations.
The low cross section of the $\Upsilon$ at RHIC energies make this a 
luminosity limited measurement.
The upcoming measurement in heavy-ion collisions will shed more light 
into the expected melting of quarkonia states in the hot and dense 
medium and provide an estimate of the medium temperature.

\section{Summary and conclusions}
In this paper, we summarize recent heavy flavor results from the STAR 
experiment at RHIC.
The charm cross-section was extracted from a combined fit to the
measured spectra of open charm mesons, electrons and muons 
both from semi-leptonic heavy flavor decays.
The charm cross section in Au+Au collisions scales with 
the number of binary collisions supporting the assumption that 
charm is exclusively produced in the initial state of the collision and 
that there is no room for thermal production in the medium.
The suppression of the nuclear modification factor of non-photonic 
electrons at high $\pt$ in Au+Au collisions is much larger than 
expected. The theoretical explanations are yet inconclusive.
First results on azimuthal angular correlations of non-photonic 
electrons and hadrons ($D^0$ mesons) in p+p collisions show its 
ability to disentangle the $D$ and $B$ contribution to the electron 
spectrum.
The recent completion of the STAR EMC allowed the first measurement 
of $\Upsilon$ production at mid-rapidity in p+p collisions.
The $\Upsilon$ cross section is consistent with pQCD calculations 
and the world data trend.
More exciting results are about to come with the STAR detector 
upgrades (full barrel Time-of-Flight and Heavy Flavor Tracker) which 
will allow direct measurements of the nuclear modification factor of 
$D$ and $B$ mesons in heavy-ion collisions.

\begin{footnotesize}

\end{footnotesize}

\end{document}